\documentstyle[11pt]{article}
 \newcommand {\nc}{\newcommand}
 \nc{\eq}{\begin{equation}}
 \nc{\en}{\end{equation}}
 \nc {\norm}[1]{\parallel{#1}\parallel}
 \def\intg{{\cal Z}}
 
 \def\complex{{\cal C}}
 \def\Alg{{\cal A}}
 \def\Hs{{\cal H}}
 
 \def\unit{{\bf 1}}
 
 \def\wtld{\widetilde}
 
 \def\dlt{\delta}
\begin{document}
  \title{Vacuum Solutions of Classical Gravity on Cyclic Groups from Noncommutative Geometry}
  \author{Jian Dai, Xing-Chang Song\\
   Theory Group, Department of Physics, Peking University\\
   Beijing, P. R. China, 100871\\
   jdai@mail.phy.pku.edu.cn,
   songxc@ibm320h.phy.pku.edu.cn}
  \date{
  February 5th, 2001\\
  Revised on April 4th, 2001}
  \maketitle
  \begin{abstract}
   Based on the observation that the moduli of a link variable on a cyclic group modify Connes' distance on this group, we
   construct several action functionals for this link variable
   within the framework of noncommutative geometry. After solving
   the equations of motion, we find that one type of action gives
   nontrivial vacuum solution for gravity on this cyclic group in a broad range of coupling
   constants and that such solutions can be expressed with
   Chebyshev's polynomials. \\


  \end{abstract}
  \section{Introduction}
   It is a marvelous property of {\it noncommutative
   geometry}(NCG) that Dirac operator induces a metric onto a
   space \cite{new}. Specifying this result to a cyclic group, we discussed recently at the end of \cite{dsW} that the effect of
   a non-unitary 1D link variable on this {\it Connes' distance} is to modify a linear(Euclidean)-distance with the moduli of
   this link variable. This feature indicates the general characteristics of NCG that
   gauge connection and metric are easily interwound on a
   noncommutative space. In this contribution, we will explore the
   (classical) gravitational physics of cyclic groups in depth.
   We construct action functionals for gravitational fields from Dirac operator, deduce the equations of
   motion, then solve them and give a series of vacuum solutions. It will
   be shown that in a broad range of coupling constants of one action functional, there is a
   non-trivial vacuum solution that can be expressed by Chebyshev's polynomials and that breaks translation
   invariance. Of course, this model is just an exercise as a class of toy-models like \cite{Rovelli}\cite{hale};
   however, it provides
   an easily-handling example to formulate gravity theory on
   discrete sets. Though being simple, it has shown a lot of general features of NCG
   approach to gravity problems. For a more abstract and general treatment of noncommutative gravity,
   one can refer to Majid's \cite{Majid1}\cite{Majid2}.\\

   This paper is organized as the following way. In section
   \ref{sectA}, NCG on a cyclic group is formulated, especially
   Dirac operator is defined. Several action functionals are
   established in section \ref{sectB}. Three types of vacua will be found out, in which only the last one is
   nontrivial. Some opening discussions are put into section
   \ref{sectC}.
  \section{NCG of Cyclic Groups: Kinematics of Dirac Operator}\label{sectA}
   Let $\intg_N$ be a $N$-order cyclic group
   $\intg_N=\{0,1,2,\ldots,N-1\}$ whose multiplication is
   just integer addition modulo $N$. Additions appearing below are
   understood as additions in $\intg_N$. $\Alg(\intg_N)$ is the algebra of complex functions
   on $\intg_N$ and there is a {\it regular representation} of $\intg_N$ on $\Alg(\intg_N)$ generated by
   $(T^+f)(x)=f(x+1), \forall f\in\Alg(\intg_N),x\in\intg_N$. For any finite $N$, $(T^+)^N=\unit$ where
   $\unit$ is identity transformation on $\Alg(\intg_N)$;
   hence $T^-:=(T^+)^{-1}=(T^+)^{N-1}$.
   A spinor space $\Hs_s=\complex^2$ is introduced to each
   point of $\intg_N$; a {\it fermion field} on $\intg_N$ is an element
   in $\Hs:=\Alg(\intg_N)\otimes\Hs_s$. Under the standard inner product on $\Hs$,
   $T^+$ is a unitary operator.
   A {\it Fredholm operator} acting on $\Hs$ is defined to be
   $F=\eta_++\eta_-$
   in which $\eta_\pm=T^\pm\sigma_\pm$,
   $\sigma_\pm=(\sigma_1\pm i\sigma_2)/2$ and $\sigma_i, i=1,2,3$ are Pauli matrices \cite{DM_D}.
   Note that $\eta_\pm$ fulfil Clifford algebra relation on 2D-Euclidean
   space $\eta_\pm\eta_\pm=0, \{\eta_\pm,\eta_\mp\}=\unit$; accordingly, $F^2=\unit$, s.t.$(\Hs,F)$ forms a {\it
   Fredholm module} \cite{Connes_Book}.
   The fundamental {\it
   non-commutativity} in this formalism is $\eta_\pm f=(T^\pm
   f)\eta_\pm$.
   It is proved in \cite{DM_D} that Connes' distance on $\intg_N$ defined
   by $d_F(x,y)=sup\{|f(x)-f(y)|:\norm{[F,f]}\leq 1\}, \forall
    x,y\in\intg_N$
   is just the conventional linear(Euclidean)-distance on
   $\intg_N$: $d(x,y)=min\{|x-y|,N-|x-y|\}$.\\

   Dirac operator on $\intg_N$ is $F$ twisted with a
   link-variable:
   $F(\omega)=\omega^\dag\eta_++\eta_-\omega$
   in which $\omega$ is a
   $End_\complex(\Hs_c)$-valued function on $\intg_N$ and $\Hs_c$ is a internal Hilbert space.
   The triple $(\Alg(\intg_N),\wtld{\Hs}, F(\omega))$ becomes
   a {\it K-cycle} where $\wtld{\Hs}=\Hs\otimes \Hs_c$ \cite{Connes_AX}.
   A gauge transformation $u$ is a $U(\Hs_c)$-valued function on
   $\intg_N$; $T^-\omega$, $\omega^\dag T^+$ play the role of
   $U(\Hs_c)$ {\it parallel transports}, providing that
   $\omega\rightarrow (T^+ u)\omega u^\dag$.
   Below only 1-dimensional case $\Hs_c=\complex$ will be considered, hence $\omega\in\Alg(\intg_N)$.
   Subsequently, $\omega=\rho e^{i\theta}$ where $\rho,\theta$ are real functions, $\rho(x)\geq 0$ and
   $\omega^\dag=\overline{\omega}$, $\omega^\dag\omega=|\omega|^2=\rho^2$.
   We argued at the end of \cite{dsW} that $d_{F(\omega)}(x,x+k)=
   min\{\rho(x)^{-1}+\rho(x+1)^{-1}+\ldots +\rho(x+k-1)^{-1},\rho(x+k)^{-1}+\rho(x+k+1)^{-1}+\ldots
   +\rho(x+N-1)^{-1}\}$
   for all $x,k\in \intg_N$. Therefore, metric on $\intg_N$ is
   modified by the strength of $\omega$ in the sense that $\rho^{-1}$ provides
   a varying lattice spacing. This point can
   be illustrated more clearly by some special examples.
   1)$\rho(x)=\rho_0 >0$, then resulting metric differs from the ``free'' one by a lattice constant
   $1/\rho_0$;
   2)$\rho(0)=0$, then $d_{F(\omega)}(0,1)=\infty$, which can be interpreted that the points $x=0$ and $x=1$ are
   disconnected;
   3)$\rho(0)\rightarrow +\infty$, then $d_{F(\omega)}(0,1)\rightarrow 0$, which is interpreted as a ``black hole'' by M. Hale in
   \cite{hale}.
   In the next section, dynamics of $\omega$ will be considered;
   cases 1) and 2) will be shown to emerge from classical
   solutions to the equations of motion for $\omega$. We deliver two identities
   at the end of this section
   \eq\label{F2}
    F(\omega)^2=T^-(\omega\omega^\dag)\eta_-\eta_+ +(\omega^\dag
    \omega)\eta_+\eta_-
    =T^-(\rho^2)\sigma_-\sigma_+ +(\rho^2)\sigma_+\sigma_-
   \en
   \eq\label{FW2}
    F(\omega)\wedge F(\omega):=T^-(\omega\omega^\dag)\eta_-\wedge\eta_+ +(\omega^\dag
    \omega)\eta_+\wedge\eta_-
    ={1\over 2}(T^-(\rho^2)-\rho^2)[\sigma_-,\sigma_+]
    =-{1\over 2}\partial^-(\rho^2)\sigma_3
   \en
   where $\partial^\pm f:=T^\pm f-f$.
  \section{Vacuum Solutions: Dynamics of Dirac Operator}\label{sectB}
   Three action functionals containing only $\omega$ will be considered
   below. What we mainly concern is whether there be non-trivial
   {\it vacuum solutions} in the corresponding equations of motion. By a
   ``non-trivial vacuum'', we mean that translation invariance of $\intg_N$ is
   broken by this solution.
   \subsection{Trivial Cases: $S[\omega]=Tr(F(\omega)^{2k}), k=1,2,...$}
    Notice Eq.(\ref{F2}),
    $S[\omega]=2\sum\rho^{2k}$. This action admits only 0-solution obviously.
   \subsection{Trivial Cases: $S[\omega]=Tr((F(\omega)\wedge F(\omega))(F(\omega)\wedge
   F(\omega)))$}
    By Eq.(\ref{FW2}), $S[\omega]={1\over
    2}\sum(\partial^-(\rho^2))^2$. Equation of motion is deduced
    by $\dlt_\omega S[\omega]=0$
   \eq\label{EQM}
    \rho(x)(2\rho(x)^2-\rho(x+1)^2-\rho(x-1)^2)=0, \forall x\in
    \intg_N
   \en
   If $\rho(x)\neq 0,\forall x\in\intg$, let $\phi=\rho^2$,
   Eq.(\ref{EQM}) takes on the form
   \eq\label{HARM}
    \partial^+\partial^- \phi=0
   \en
   which is a discretized harmonic equation $\Delta\phi=0$ on $S^1$. Eq.(\ref{HARM}) admits only constant solution
   $\rho(x)=\rho_0$, which can be understood schematically by the {\it extremal value principle} in commutative harmonic
   analysis \cite{MPEQ}. The {\it on-shell} metric satisfies $d_{F(\omega)}(,)={1\over
   \rho_0}d_F(,)$ which corresponds the case 1) in the last
   section, namely that lattice constant is modified from 1 to
   $1/\rho_0$.
   Else if there is one $x_0$ such that $\rho(x_0)=0$, then the
   only consistent solution is $\rho(x)=0$ for all $x$.
  \subsection{Nontrivial Case}
   Now consider
   \eq\label{ntact}
    S[\omega]={1\over 2}Tr((F(\omega)\wedge F(\omega))(F(\omega)\wedge F(\omega))
    +{\alpha\over 4}F(\omega)^4-{\beta\over 2}F(\omega)^2)
   \en
   in which coupling constants $\alpha,\beta$ are real numbers and are required to be not equal to zero at
   the same time. One can check that $S[\omega]=
   \sum({1\over 4}(\partial^-(\rho^2))^2+{\alpha\over 4}\rho^4-{\beta\over
   2}\rho^2)$. The equation of motion is given by
   \eq\label{NTEQ}
    \rho(2t\rho^2-T^+(\rho^2)-T^-(\rho^2)-\beta)=0
   \en
   where $t:=(2+\alpha)/2$.
  \subsubsection{Non-singular Case: $\rho(x)>0, \forall
  x\in\intg_N$}
   Remember the definition of $\phi$, then Eq.(\ref{NTEQ}) changes
   form as
   \eq\label{eC}
    -\phi(x-1)+2t\phi(x)-\phi(x+1)=\beta, \forall x\in\intg_N
   \en
   The cyclic symmetry in Eq.(\ref{eC}) $\phi(x)\rightarrow \phi(x+1)$ implies that $\phi(x)=\phi_0, \forall
   x\in\intg_N$ in which $\phi_0={\beta\over{2t-2}}={\beta\over
   \alpha}$. Note that if $\alpha=0,\beta\neq 0$, no solution
   exists. Upon this solution as a background, metric fulfills relation
   $d_{F(\omega)}(,)=\sqrt{\alpha\over\beta}d_F(,)$.
  \subsubsection{Singular Case}\label{sing}
   The remainder of this section will be devoted to one most
   interesting situation in which, without losing generality, $\rho(N-1)=0, \rho(x)>0, x=0,1,\ldots N-2$.
   If we find out a solution to this case, then this solution breaks translation invariance, hence being a nontrivial vacuum.
   Some matrix algebra has to be prepared here. Introduce a matrix sequence
   $\{M_n(t):n=1,2,...\}$ in which
   \[
    M_n(t):=\left(
     \begin{array}{cccccccc}
      2t&-1&0&0&\ldots&0&0&0\\
      -1&2t&-1&0&\ldots&0&0&0\\
      0&-1&2t&-1&\ldots&0&0&0\\
      &&&&\ddots&&&\\
      0&0&0&0&\ldots&-1&2t&-1\\
      0&0&0&0&\ldots&0&-1&2t
     \end{array}\right)
   \]
   is a n$\times$n matrix, and define
   \eq\label{def}
    U_n(t)=det(M_n(t))
   \en
   One can prove this iterative relation easily
   \eq\label{it}
    U_0(t):=1,U_1(t)=t, U_n(t)=2tU_{n-1}(t)-U_{n-2}(t), n=2,3,...
   \en
   Eq.(\ref{it}) shows that $U_n(t)$ are {\it Chebyshev's polynomials
   of the second kind}, hence
   $U_n(t)=\frac{z_+^{n+1}-z_-^{n+1}}{2\sqrt{t^2-1}}$
   in which $z_\pm=t\pm \sqrt{t^2-1}$ \cite{Table}. So (\ref{def})
   provides Chebyshev's polynomials with another interpretation.
   Note that all roots of $U_n(t)$ are real and lie in $(-1,1)$
   and that $U_n(t)>0$, if $t>1$, $n=0,1,2,\ldots$.
   We set $t>1$, i.e. $\alpha>0$ from now on.\\

   Now let $\Phi=(\phi(0),\phi(1),\ldots,\phi(N-2))^T$,
   $\unit_V={\underbrace{(1,1,\ldots,1)}_{N-1}}^T$, then equation of motion (\ref{NTEQ})
   can be written as $M_{N-1}(t)\Phi=\beta\unit_V$.
   Assume $\beta\neq 0$ and scale $\phi=\beta v, \Phi=\beta V$,
   then
   \eq\label{Eequation}
    M_{N-1}(t)V=\unit_V
   \en
   Introduce notations $f_{i,j}(t)=\frac{U_i(t)}{U_j(t)}$,
   $\Sigma_n(t)=\sum_{j=0}^nU_j(t)$,
   $\psi_n=\frac{\Sigma_n(t)}{U_{n+1}(t)}$. Formal solution to Eq.(\ref{Eequation}) is
   \eq\label{NTsolution}
    v(x)=\sum_{y=0}^{N-2-x}f_{x,x+y}(t)\psi_{x+y}(t),
    x=0,1,...,N-2
   \en
   Since we choose $t>1$, there is no singularity in $v(x)$.
   If $\beta>0$, due to the positivity of $U_n(t)$ when $t>1(\alpha>0)$, $\phi(x)>0, \forall x=0,1,...,N-2$, i.e.
   being physically acceptable.
   Solutions for $N=3,4,5,6$ are listed in Appendix \ref{app2}.
  \section{Discussions}\label{sectC}
   The geometric interpretation of the solutions in Subsect.\ref{sing} is that when coupling
   $\alpha>0$ and $\beta>0$, there are nontrivial vacuum appearing and that the metrics appear to be
   \[d_{F(\omega)}(x,y)=\beta^{-{1\over 2}}(v(x)^{-{1\over 2}}+v(x+1)^{-{1\over 2}}+\ldots+v(y-1)^{-{1\over
   2}})\]
   for all $x,y=0,1,...,N-2,x<y$.
   Moreover, this
   statement is valid in the limit $N\rightarrow \infty$.\\

   Interesting topics along this line are interaction between link variable $\omega$ and
   matter fields, ``black hole'' solutions, quantization, $dim(\Hs_c)>1$ case and higher dimension cases. Work on these aspects is in proceeding.\\

  {\bf Acknowledgements}\\
    This work was supported by Climb-Up (Pan Deng) Project of
    Department of Science and Technology in China, Chinese
    National Science Foundation and Doctoral Programme Foundation
    of Institution of Higher Education in China.
    We are very grateful to Dr. B. Feng in MIT for that the general
    method to discuss Eq.(\ref{it}) is showed by Dr. Feng to us, and
    to Prof. S. Majid
    for introducing his work to us.
  \appendix
  \section{Solutions to Eq.(\ref{Eequation}) for N=3,4,5,6}\label{app2}
   \[N=3: v(0)=v(1)=\frac{1}{2t-1}
   \]
   \[N=4: v(0)=v(2)=\frac{2t+1}{4t^2-2}; v(1)=\frac{t+1}{2t^2-1}
   \]
   \[N=5:
    v(0)=v(3)=\frac{2t}{4t^2-2t-1}; v(1)=v(2)=\frac{2t+1}{4t^2-2t-1}
   \]
   \[N=6:
    v(0)=v(4)=\frac{4t^2+2t-1}{2t(4t^2-3)};
    v(1)=v(3)=\frac{2t+2}{4t^2-3};
    v(2)=\frac{4t^2+4t+1}{2t(4t^2-3)}
   \]
  
 \end{document}